\documentclass[,draft]{aipproc}
\layoutstyle{6x9}

\usepackage{natbib}
\usepackage{epsfig}
\usepackage{subfigure}

\begin{document}

\title{Gravitational Lensing of Neutrino from Collapsars}


\author{Florencia L. Vieyro}{
  address={Instituto Argentino de Radioastronomía (CCT- La Plata, CONICET), C.C.5, 1894 Villa Elisa, Buenos Aires, Argentina}
  ,altaddress={Facultad de Ciencias Astronómicas y Geofísicas, Universidad Nacional de La Plata, paseo del Bosque, 1900, La Plata, Argentina.}
}

\author{Gustavo E. Romero}{
  address={Instituto Argentino de Radioastronomía (CCT- La Plata, CONICET), C.C.5, 1894 Villa Elisa, Buenos Aires, Argentina}
  ,altaddress={Facultad de Ciencias Astronómicas y Geofísicas, Universidad Nacional de La Plata, paseo del Bosque, 1900, La Plata, Argentina.}
}

\begin{abstract}
 We study neutrino emission from long gamma-ray bursts. The collapse of very massive stars to black holes, and the consequent jet formation, are expected to produce high-energy neutrinos through photomeson production. Such neutrinos can escape from the source and travel up to the Earth. 
 
 We focus on the case of Population III progenitors for gamma-ray bursts. Neutrinos can be the only source of information of the first stars formed in the universe. The expected signal is rather weak, but we propose that gravitational lensing by nearby supermassive black holes might enhance the neutrino emission in some cases. We implement a Monte Carlo analysis to ponder the statistical significance of this scenario. We suggest that an observational strategy based on gravitational lensing could lead to the detection of neutrinos from the re-ionization era of the universe with current instrumentation.
\end{abstract}

\maketitle


\section{Introduction}

  Gamma-ray bursts (GRBs) are the most violent and energetic explosive events in the universe. Since GRBs are extragalactic distant sources, the equivalent isotropic energy can be as high as $10^{51} - 10^{54}$ erg \citep{bloom2001}. The general view is that GRBs occur when an ultra-relativistic energy flow (in the form of kinetic energy of relativistic particles) is converted to internal energy through shocks and then radiated away.
  
  There are two types of GRBs: short (duration $\sim 0.1$ s) and long ($\sim 10$ s or more). This bimodality is thought to be the result of different engines in each case. Short GRBs seem to be the product of the final merger of two compact objects, whereas long GRBs are likely associated with the final collapse of very massive magnetized stars into black holes. These stars are usually called \textit{collapsars}. 
  
  In the collapsar model \citep{woosley1993} a black hole is formed in the collapsing core of a massive star. The hole accretes material from the interior of the star, generating a magnetized and hyperdense accretion disk. The magnetic field attached to the disk collimates an ultrarelativistic bipolar jet moving through the star. The jet pushes the stellar material outwards. When the relativistic jet escapes form the star, strong shocks are expected to convert bulk motion into relativistic particles. These relativistic particles interact with the magnetic, matter and photon fields, producing gamma rays when the jet emerges from the star (uncorking process, see \citep{woosley1993,harikae}). In this process, in addition to the gamma-rays, photomeson production can lead to neutrino emission. The neutrino signal, however, is likely to be rather weak, making the detection with current instruments difficult.

In this work, we propose that gravitational lensing by nearby supermassive black holes can enhance the neutrino emission allowing for detections.

\section{Neutrino production}

Charged pions and muons are injected in the jet through hadronic interactions in the uncorking region. These transient particles decay
producing neutrinos:

	\begin{equation}
		 \pi^{\pm} \rightarrow \mu^{\pm} + \nu_{\mu}(\overline{\nu}_{\mu}) \textrm{,}
	\end{equation}
	
	\begin{equation}
		 \mu^{\pm} \rightarrow e^{\pm} + \overline{\nu}_{\mu}(\nu_{\mu}) + \nu_{e}(\overline{\nu}_{e}) .
	\end{equation}

\begin{figure}[h]
	\includegraphics[width=1.0\textwidth, keepaspectratio]{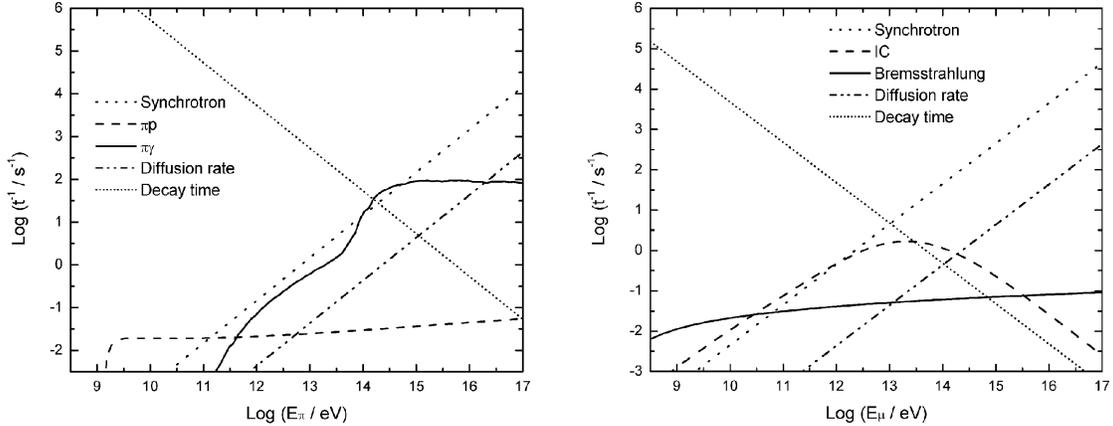}
  \caption{Radiative losses of charged pions and muons in a magnetized plasma located in the uncorking region of a collapsar. Stellar radius and mass are $\sim 10^{12}$ cm and $\sim 100 M_{\odot}$, respectively.}
  \label{fig:perdidas}
\end{figure}

At the surface of the star, relativistic particles are cooled by different mechanisms. In Fig. \ref{fig:perdidas} we show the radiative losses and decay time for charged pions and muons in a magnetized plasma ($B \sim 10^{6}$ G). From these plots it is clear that all muons with energies above $E_{\mu} \sim 10^{13}$ eV are cooled with the consequent neutrino attenuation. For pions, suppression of high energy neutrinos occur above $\sim 10^{14}$ eV. As pointed out by \citep{reynoso2009} these effects lead to a significant neutrino signal attenuation in instruments such as IceCube, sensitive only to high energy neutrinos ($E_{\nu} > 1$ TeV).


\section{Gravitational lensing}

Neutrinos produced in a GRB \citep{meszaros2001} can escape from the source and travel up to the Earth. It is of special interest to study the case of Population III progenitors for the gamma-ray bursts \citep{gao2011,berezinsky2011}, since the neutrinos can be the only source of information of these stars.

\begin{figure}[h]
	\includegraphics[width=0.7\textwidth, keepaspectratio]{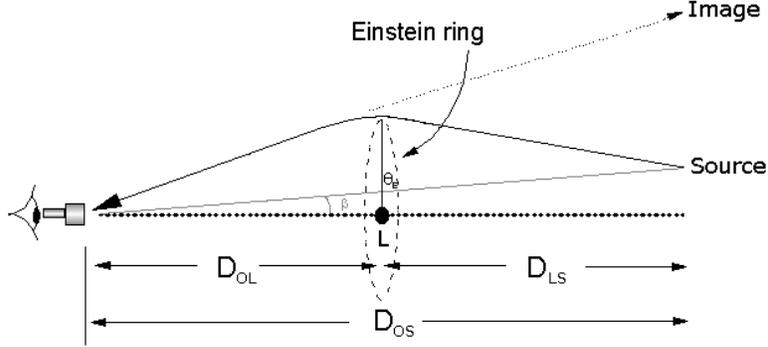}
  \caption{Lens diagram.}
  \label{fig:lens}
\end{figure}

Gravitational lensing is the phenomenon resulting from the deflection of a signal in a gravitational field \citep{einstein}. We study the magnification of the neutrino signal by gravitational lensing by nearby supermassive black holes. Neutrinos have zero or negligible mass, so they follow in very good approximation null geodesics \citep{eiroa}. We consider a point source of neutrinos, with distance $D_{\rm{os}}$ to the observer, behind a Schwarzschild black hole lens, placed at a distance $D_{\rm{ol}}$. The distance between the lens and the source is $D_{\rm{ls}}$ (see Fig. \ref{fig:lens}).

The angular position of the source $\beta$ is small when the alignment is high. For this configuration, two weak deflection images and two infinite sets of strong deflection images (also called relativistic) are formed \citep{virbhadra}. Typical distances between the images in gravitational lensing are of the order of the Einstein radius, given by:

\begin{equation}
	\theta_{\rm{E}}= \sqrt{\frac{2R_{\rm{Schw}}D_{\rm{ls}}}{D_{\rm{ol}}D_{\rm{os}}}} ,
\end{equation}

\noindent where $R_{\rm{Schw}}$ is the Schwarzschild radius of the lens. When $\beta=0$, instead of two images, an Einstein ring with radius $\theta_{\rm{E}}$ is obtained. 

An important aspect is the magnification of the images, defined as the ratio between the observed and intrinsic fluxes of the source. As a consequence of the Liouville theorem in curved spacetimes, gravitational lensing preserves surface brightness for neutrinos and photons, so the magnifications of the images are given by the ratio of the solid angles subtended by the images and the source. In ref. \citep{romeroVieyro2011}, it is shown that the gravitational lensing produced by a supermassive black hole as the one in the center of the galaxy NGC4486 (M87), can increase the total number of neutrinos at Earth by an order of magnitude.

\section{Statistical analysis}

\begin{figure}
	\includegraphics[width=0.7\textwidth, keepaspectratio,angle=270]{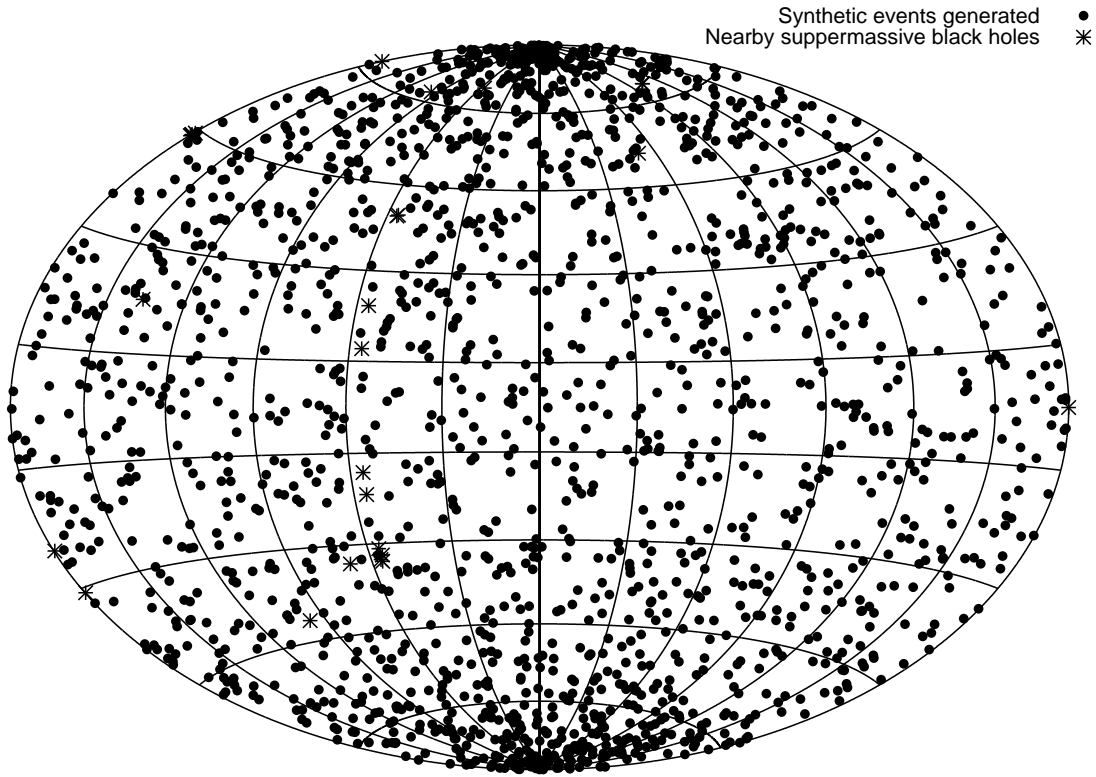}
  \caption{Simulated five-year, all-sky GRBs map in Hammer-Aitoff projection (Galactic coordinates). The position of the nearby supermassive black holes of the sample are indicated with a star; the black circles indicate the positions of the events synthetically generated for 5 years at a rate of 1 long GRB per day.}
  \label{fig:mapaSources}
\end{figure}

Using a Monte Carlo method, we study the possibility that a GRB can be detected inside the Einstein rings of a group of nearby galaxies with known supermassive black holes. We use a sample of 20 galaxies of the catalog of nearby black holes candidates \citep{caramete} and we also include the galaxies discussed in \citep{eiroa}. As an example, Table \ref{table} lists ten galaxies of the sample, the value of their main parameters, including the estimated Einstein ring radius. We generate synthetic events uniformly distributed in the sky through a Monte Carlo routine, and count the events that are located inside the Einstein radius of a supermassive black hole (see Fig. \ref{fig:mapaSources}). 

\begin{table}[h]
\centering
\begin{tabular}{l | c | c | c}
\hline  
Galaxy & Distance [Mpc]$^{(1)}$ & Black holes mass [$10^{8}M_{\odot}$]$^{(1)}$  & $\theta_{\rm{E}}$ [arcsec]  \\[0.4cm]
\hline 

Milky Way      	 & $0.0085$      & $0.03$     & $1.64$   \\[0.01cm]
NGC0224          & $0.7$      	 & $0.3$      & $0.59$    \\[0.01cm]
MESSIER094  	   & $4.6$ 				 & $0.75$     & $0.36$     \\[0.01cm]      
NGC3115       	 & $8.4$			   & $20.0$  	  & $1.39$  \\[0.01cm]
NGC4486  	     	& $15.3$ 	    	 & $33.0$   	& $1.32$   \\[0.01cm]
NGC5846 		  	& $29.1$ 				 & $5.37$     & $0.38$    \\[0.01cm]   
NGC5850 				& $35.0$ 			   & $2.68$     & $0.25$   \\[0.01cm]
NGC6500 				& $41.1$ 				 & $1.41$     & $0.17$   \\[0.01cm]
NGC7469 				& $67.0$ 				 & $4.96$     & $0.27$  \\[0.01cm]
UGC00600 				& $93.3$ 				 & $2.06$     & $0.13$   \\[0.01cm]

\hline\\[0.005cm]
\multicolumn{2}{l}{
$^{(1)}$ Values from \citep{eiroa}and \citep{caramete}}				  \\[0.01cm]
\end{tabular}	
	\caption{Parameters of ten galaxies with supermassive black holes of the sample.}
  \label{table}
\end{table} 

The GRB detection rate for the BATSE-CGRO detector was about one per day during its operation period 1991-2000 \citep{cohen1995}. Many collapsars can fail to produce a $\gamma$-ray signal but nonetheless still they can be $\nu$-sources. We assume a rate of one long GRB event per day. The current instrument for detecting GRBs, the Burst Alert Telescope (BAT), locates the position of each event with an accuracy of 1 to 4 arc-minutes, hence we represent each generated event as a circle with a radius of 2 arcmin centered in the galactic coordinates $(\alpha,\beta)$, randomly generated.

After generating a significant number of events, the estimated frequency of detecting with full IceCube a $\nu$-GRB affected by gravitational lensing results in $\sim 1$ event in 5 years.

Despite the low probability that gravitational lensing can enhance the neutrino emission from GRBs, the observational strategy proposed in this work is a tool that could be useful for cosmology and theory of structure formation, since it can provide unique information on neutrinos from the re-ionization era.

\begin{theacknowledgments}
  This work was partially supported by the Argentine agencies ANPCyT (BID 1728/OC -AR PICT-2007-00848) and CONICET (PIP 0078/2010).
\end{theacknowledgments}

\end{document}